\documentclass[aps,prl,twocolumn,superscriptaddress,showpacs]{revtex4}

\usepackage{graphicx}

\def\scgw{{sc{\em GW}}}

\def\GLDA{{G^{\rm LDA}}}
\def\WLDA{{W^{\rm LDA}}}

\begin{document}


\title{
All-electron self-consistent GW approximation: Application to Si, MnO, and NiO
}



\author{Sergey V. Faleev}
\affiliation{Sandia National Laboratories, Livermore, CA 94551}

\author{Mark van Schilfgaarde}
\affiliation{Arizona State University, Tempe, AZ, 85284}

\author{Takao Kotani}
\affiliation{Department of Physics, Osaka University, Toyonaka 560, Japan}

\date{\today}


\begin{abstract}
We present a new kind self-consistent {\em GW} approximation
(\scgw) based on the all-electron, full-potential LMTO method.
By iterating the eigenfunctions of the $GW$ Hamiltonian,
self-consistency in both the charge density and the quasiparticle
spectrum is achieved.  We explain why this form of
self-consistency should be preferred to the conventional one.
Then some results for Si are shown as a representative
semiconductor, to establish agreement with a prior \scgw\
calculation.  Finally we consider many details in the electronic
structure of the antiferromagnetic insulators MnO and NiO.
Excellent agreement with experiment is shown for many properties,
suggesting that a Landau quasiparticle (energy band) picture
of MnO and NiO provides a reasonable description of electronic
structure even in these correlated materials.
\end{abstract}


\maketitle

The $GW$ approximation (GWA) of Hedin\cite{hedin65} is generally
believed to accurately predict excited-state properties, and in
particular improve on the local density approximation (LDA),
whose limitations are well known, e.g. to underestimate bandgaps
semiconductors and insulators.  Usually GWA is computed as
1-shot calculation starting from the LDA eigenfunctions and
eigenvalues; the self-energy $\Sigma$ is approximated as
$\Sigma= i\GLDA\WLDA$, where $\GLDA$ is a bare Green function
constructed from LDA eigenfunctions, and $\WLDA$ is the screened
Coulomb interaction constructed from $\GLDA$ in the random phase
approximation (RPA).  However, establishing the validity of the
1-shot approach has been seriously hampered by the fact that
nearly all calculations to date make further approximations,
e.g. computing $\Sigma$ from valence electrons only; the
plasmon-pole approximations; and the pseudopotential (PP)
approximation to deal with the core.  Only recently when reliable
all-electron implementations have begun to appear, has it been
shown that the 1-shot GWA with PP leads to systematic
errors\cite{kotani02,alouani03,weiku02}.  There is general
agreement among the all-electron calculations (see
Table~\ref{tab:si}) that the $\Gamma$-X transition in Si is
underestimated when $\Sigma=i \GLDA\WLDA$.  And we have
shown previously\cite{kotani02} that the tendency for
$\Sigma= i \GLDA \WLDA$ to underestimate gaps is almost universal in
semiconductors.  This is reasonable because $\WLDA$
overestimates the screening owing to the LDA small band gaps.
$G$ constructed from quasiparticles (QP) with a wider gap (e.g.\ a self-consistent
$G$) reduces the screening, and therefore generates $GW$ with a wider gap. 

  However, there are many possible ways to achieve
self-consistency.  The theoretically simplest (and internally
consistent) is the fully self-consistent scheme (\scgw), which is
derived through the Luttinger-Ward functional with the
exchange-correlation energy approximated as the sum of RPA ring
diagrams.  Then $W$ is evaluated as $W = v(1-v P)^{-1}$ with the
proper part of the polarization function $P\!=\!-iG \times G$.
However, such a construction may not give reasonable $W$
\cite{tamme99}, resulting in a poor $G$, for the following
reason.  If $\Sigma$ is $\omega$-dependent, $G$ can be
partitioned into a QP part and a residual satellite part.  The QP
part consists of terms whose energy-dependence varies as
$Z_i/(\omega\!-\epsilon_i\!\pm\!i\Gamma_i)$, where $\epsilon_i$,
$\Gamma_i$, and $Z_i$ are respectively the QP energies, inverse
lifetimes, and renormalization factors ($Z_i${}$<$1; typically
between 0.7 and 1).  The QP parts are thus weighted by factors
$Z$; the residual weights $1\!-\!Z$ go into the plasmon-related
satellite parts, high in energy.  Thus $P\!=\!-iG \!\times\! G$
contains contributions from the particle-hole pair excitations as
does $-i\GLDA\!\times\!\GLDA$, but reduced by the products of two
$Z$ factors, one from occupied and the other from unoccupied
states.  However, this construction of $P$ is not consistent with
Landau's quasi-particle theory, which insists that one-particle
excitations remain meaningful (at least near the Fermi energy).
Based on the theory, we should instead evaluate the QP
contributions to $P$ without $Z$ factors, as they dominate the
static screening $W(\omega\!\to\!0)$.  Inclusion of $Z$ can lead
to $W(0)$ being underscreened; moreover $W(\omega)$ does not
satisfy the $f$ sum rule \cite{tamme99}.  Consequently $W(0)$
will be overestimated resulting in a tendency to overestimate
band widths, as is well known in the extreme case (Hartree-Fock),
where there is no screening of $W$.

  Indeed Holm found that \scgw\
overestimates the band width in the
homogeneous electron gas\cite{holm98}.  Very recently Ku
presented a \scgw\
calculation which similarly overestimates the valence band width
in Si and Ge; see Table~\ref{tab:si} \cite{weiku02}.
Another practical justification for the argument that a bare $G$ without $Z$
should be used when we construct $P$ as $-iG\times{}G$,
is that $\WLDA$ is already known to be rather good
if we add some enlargement of band gap by hand to correct for
errors in the LDA $\epsilon_i$ \cite{arnaud01}.

For this reason, we do not adopt the full \scgw\ scheme, but
construct two kinds of constrained self-consistent $GW$
methods.  For a set of trial eigenfunctions
and quasi-particle energies $\{\psi_{{\bf q}n},\epsilon_{{\bf q}n} \}$,
we can calculate the one-particle Green function $G$,
from which we can calculate self-energy
$\Sigma^{{{\bf{q}}}}_{nn'}(\omega)$ in the expansion of
$\{\psi_{{\bf q}n}\}$ in the GWA.  Then we generate an
energy-independent and hermitian $\Sigma_{nn'}^{^{{\bf{q}}} }$ in
one of two ways:
\begin{eqnarray}
\Sigma_{nn'}^{^{{\bf{q}}} } &=&
 {\Sigma _{nn'}^{^{{\bf{q}}} } }\left( { E_{\rm F} } \right) + \delta_{nn'}{\rm Re}[\Sigma^{{\bf q}}_{nn}(\varepsilon_{{\bf q}n}) - \Sigma^{{\bf q}}_
{nn}(E_{\rm F})]\label{mode1}\quad \\
\Sigma _{nn'}^{^{{\bf{q}}} } &=&
   {{{\rm Re}\left[ {\Sigma _{nn'}^{^{{\bf{q}}} } \left( {\varepsilon _{{{\bf{q}}n} } } \right) + \Sigma _{nn'}^{^{{\bf{q}}} } \left( {\varepsilon
_{{{\bf {q}}n'} } } \right)} \right]}/{2}}
\label{mode2}
\end{eqnarray}
where Re means that we take only the hermitian parts.
With this $\Sigma_{nn'}^{^{{\bf{q}}}}$, we can construct a new
density $n(\bf{r})$ and corresponding Hartree potential, and
proceed to a new set of
$\{\psi_{{\bf{}q}n},\epsilon_{{\bf{}q}n}\}$.  The procedure
can be iterated to self-consistency.  Our method is not related to
the LDA (though in practice the LDA is used to make a starting
guess for $\Sigma$ and the augmented-wave basis set).  These two
schemes differ in the treatment of the off-diagonal parts of
$\Sigma_{nn'}$; but both restrict the potential to be
non-local, hermitian and $\omega$-independent.  Thus the problem in the
full \scgw\ is avoided; also the numerical computation becomes
rather stable.  In Ref.\cite{weiku02}, off-diagonal matrix
elements of $\Sigma_{nn'}^{^{{\bf{q}}} }$ were completely
neglected.  However, these are important for MnO and NiO:
eigenfunctions and the density can not be changed from LDA if we neglect
them.  We find that converged QP energies differ in these two
schemes by small amounts (less then $<0.02$~eV for Si, typically
$\sim0.1$~eV for NiO), which is within the resolution of the
method ($\sim$0.1~eV).

Our implementation is based on the method of Ref.~\cite{kotani02}.
$W$ is expanded in a mixed basis which consists of two
contributions, local atom-centered functions (product basis)
confined to muffin-tin spheres, and plane waves with the
overlap to the local functions projected out.  The former can
include any of the core states: thus the valence and core states
can be treated on an equal footing and the contribution of the
latter to $\Sigma$ included.  We calculate the full energy
dependence of $W$ without the plasmon-pole approximation.  This
approach shares some features in common with both the
full-potential, all-electron plane-wave based
methods\cite{alouani03,weiku02} and the product-basis
method\cite{ferdi98}, combining the advantages of each, e.g.
efficient treatment of localized valence electrons.

Results for Si are shown in Table~\ref{tab:si}.  Agreement
between the three all-electron methods is generally excellent.
The $\GLDA\WLDA$ gaps are $\sim$0.3~eV smaller than
experiment; the \scgw\ gaps fall much closer.  As we will show
elsewhere, most properties of weakly correlated systems
calculated with the present \scgw\ method (fundamental and
higher-lying gaps, valence bandwidths, effective mass, position
of deep $d$ levels) are in excellent agreement with experiment,
with small but systematic residual errors.

\begin{table}
\caption{
Minimum energy gap $E_g$ and selected energy eigenvalues for Si,
relative to $\Gamma'_{25v}$ (eV).  Three all-electron methods are
shown: linearized augmented-plane-wave (LAPW) and
projector-augmented-wave (PAW) approaches, and LMTO (this
work). The PAW calculation included valence electrons only.  The
last row compares the Ge valence bandwidth.  The results of this
work differ slightly from Ref.~\cite{kotani02} because a large
basis set (50 orbitals/atom) was employed in the present work.
\label{tab:si}
}
\begin{tabular}{lccccccl}
                &PAW\cite{alouani03}  &\multicolumn{2}{c}{LAPW\cite{weiku02}}  & \multicolumn{2}{c}{This work}   & Exp.  \\
                &$(GW)^{\rm LDA}$
                  & $(GW)^{\rm LDA}$ &\scgw
                                   & $(GW)^{\rm LDA}$ &  \scgw &    \\
\colrule
$E_g$          & 0.92  &0.85 & 1.03 &0.92 &1.14 &1.17   \\
$X_{1c}$ &       1.01  &     &      &1.06 &1.30 &1.32   \\
$L_{1c}$ &       2.05  &     &      &2.00 &2.26 &2.04   \\
$\Gamma_{15c}$ & 3.09  &3.12 & 3.48 &3.11 &3.40 &3.38   \\
$\Gamma_{1v}$ &        &-12.1&-13.5&-12.1 &-12.3 &-12.5 \\
$\Gamma_{1v}$\rlap{(Ge)}&&-13.1&-14.8&-12.9 &-13.1 &-12.6
\end{tabular}
\end{table}

\begin{figure}[ht]
\centering
\includegraphics[width=8.5cm]{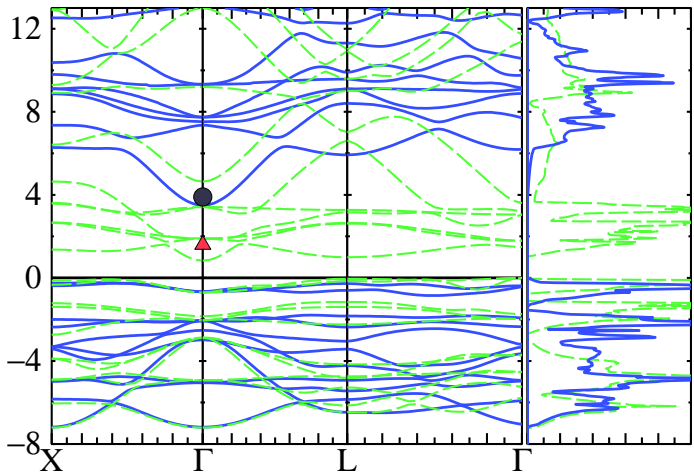}
\caption{\scgw\  energy bands and DOS of MnO.  Solid lines: \scgw\
  bands; dashed lines: LDA bands.  The VBM is set to energy 0.
  Circle and triangle at $\Gamma$: BIS and $\GLDA\WLDA$ gaps.
  Right panel shows the corresponding DOS.  Peaks
  at -0.5~eV (-0.1~eV in LDA) and -5~eV are the nearly
  dispersionless $e_{g}$ bands.  Peaks at -2.2~eV (-1.2~eV)
  and 6.6 and 7.3~eV (1.7 and 1.9~eV) derive from
  Mn $t_{2g}$ states.}
\label{fig:bands-mno}
\end{figure}

Turning to the TM oxides, we first consider MnO because it is
less correlated.  Fig.~\ref{fig:bands-mno} compares the \scgw\
energy bands and corresponding DOS to the LDA and the
$\GLDA\WLDA$ gap.  The conduction band at $\Gamma$ is evidently a
dispersive band of $sp$ character.
Above this, fall the $t_{2g}$ bands ($\sim$6-9~eV); still higher
at $\sim$10~eV is a narrow $e_{g}$ band, whose width is
$\sim$3~eV.  Thus, the itinerant and $d$ bands are well
separated.  The minimum gap is 3.5~eV, in good agreement with
the BIS gap\cite{vanelp91} (3.9$\pm$0.4~eV).  The BIS spectrum
also shows a peak at $\sim$6.8~eV, which probably corresponds to
a convolution of the peaks of $t_{2g}$ symmetry seen in the DOS
at 6.6~eV and 7.3~eV.  These bands are in stark contrast to the
LDA, which shows the $t_{2g}$ and $e_{g}$ bands overlapping and
hybridizing with the $sp$ band at 1 to 4~eV.

LDA and \scgw\ valence bands are more similar
(Fig.~\ref{fig:bands-mno}).  In the LDA there is a narrow upper
$e_{g}$ band at 0.1~eV below the valence band maximum (VBM), and
another one at VBM$-$5~eV.  Both weakly hybridize with the O $2p$
band.  The \scgw\  pushes the upper $e_{g}$ band down to
VBM$-$0.5~eV, so that the VBM takes more O $2p$ character, and
the band at VBM$-$5~eV takes more Mn $d$ character.  The
splitting $\Delta_{v}$ between the upper $e_{g}$ level and the
$t_{2g}$ level widens from 1.0~eV(LDA) to 1.7~eV(\scgw), in good
agreement a photoemission measurement of 1.9~eV\cite{vanelp91}.
An approximately similar picture emerges from a model $GW$
calculation of Massidda\cite{Massidda95}, the most important
difference being that the model $GW$ $d$ conduction bands fall
$\sim$1~eV lower than ours.

\begin{figure}[ht]
\centering
\includegraphics[width=8.5cm]{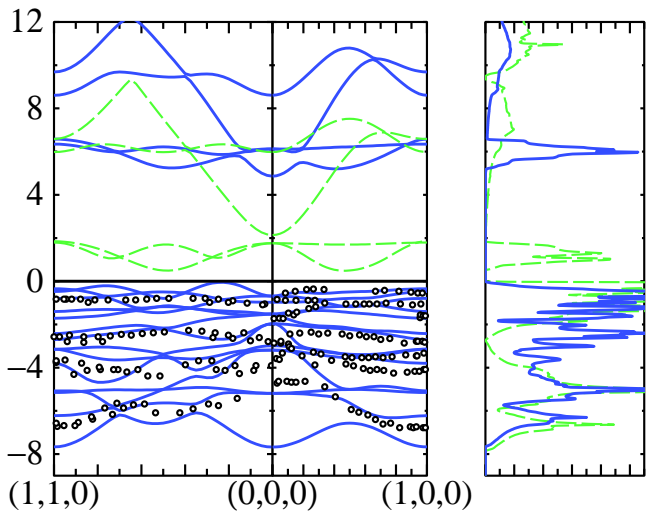}
\caption{\scgw\  energy bands and DOS of NiO.  Valence-band maximum
  is set at energy 0.  Solid lines: \scgw\  bands; dashed lines: LDA
  bands (only conduction bands are shown).  Circles show
  photoemission data of Ref.~\onlinecite{Shen}.  Right panel
  shows the corresponding total DOS.}
\label{fig:bands-nio}
\end{figure}

\begin{figure}[ht]
\centering
\includegraphics[width=8.5cm]{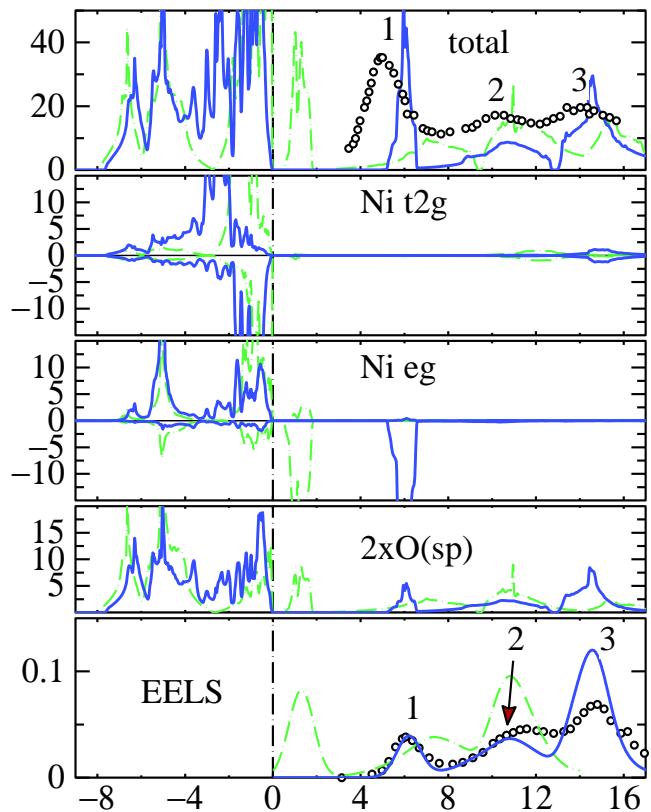}
\caption{DOS and EELS of NiO.  Solid lines: \scgw\  data; dashed
  lines: LDA data.  Top panel: total DOS, together with BIS data
  of Ref.~\onlinecite{sawatzky84} (circles).  Panels 2 and 3 show
  the Ni $t_{2g}$ and $e_{g}$ partial DOS, with positive DOS
  showing majority spin and negative showing minority spin.
  Panel 4 shows the O $sp$ partial DOS; panel 5 compares the calculated
  and measured\cite{Dudarev98} EELS spectra from the O $1s$ level.}
\label{fig:dos-nio}
\end{figure}

Fig.~\ref{fig:bands-nio} compares the \scgw\ energy bands for NiO
along the [110] and [100] lines to ARPES data of Shen \cite{Shen}
for the valence bands, and to the LDA conduction bands.  The
right panel shows the density-of-states (DOS) for both LDA and
\scgw\ , and Fig.~\ref{fig:dos-nio} shows the total DOS resolved
into components.  Also shown in the top panel are BIS data taken
from Ref.~\onlinecite{sawatzky84}.

Several features are of interest: \\
1.   The conduction-band minimum falls at the $\Gamma$ point;
       the VBM falls at the point (1/2,1/2,1/2) (not shown).
       The calculated minimum gap is 4.8~eV. \\
2.   The \scgw\  conduction bands are a mixture of a dispersive
      band composed of $sp$ approximately equally weighted on the
      Ni and O sites, and a nearly dispersionless $e_{g}$ state
      (see discussion of EELS, below).  Peaks in the BIS spectrum
      labeled `1', `2' and `3' closely coincide to those in
      the \scgw\  total DOS, apart from a constant shift of 0.8eV. \\
3.   The \scgw\  valence bands are in very good agreement with
      experiment: indeed they agree as well with the Shen data
      as the latter agrees with an independent experiment by
      Kuhlenbeck et. al.\cite{Kuhlenbeck} (not shown). \\
4.   There is an increased dispersion in the valence bands
      relative to the LDA at the VBM because the nearly
      dispersionless Ni $t_{2g}$ levels are pushed down.  Thus
      the VBM acquires somewhat more O $2p$ character.  This
      supports the generally accepted picture that the LDA too
      heavily favors the Mott-Hubbard picture.

%


The bottom panel of Fig.~\ref{fig:dos-nio} compares the EELS
spectrum calculated as described in Ref.~\onlinecite{EELS2000},
computing the excitation from the O $1s$ core.  Calculated data
were convolved by a Gaussian of 0.5~eV width (which was the
resolution reported) to compare to experimental data by reported
by Dudarev et. al.\cite{Dudarev98}, The calculated results were
shifted to align the spectra with the DOS.  For purposes of
comparison, Dudarev's data was shifted by 526 eV to align the
peaks with the \scgw\  results.  (Peaks labeled `1', `2' and `3'
should correspond to the peaks with the same labels in the BIS
spectrum; indeed the EELS peaks and BIS peaks almost perfectly
align if the EELS data is further shifted by 0.8 eV.)

Apart from the 0.8~eV shift, the EELS data is in excellent
agreement with the \scgw\ results.  Spacings between the three
peaks agree to within $\sim$0.1~eV, and the spectral weight under
each peak (estimated by numerical integration) also agree
closely. This establishes that the \scgw\ relative positions of
the {\em sp} and Ni $d$ bands are correctly predicted.  This is a
significant result, because the relative positions of the {\em
sp} and $d$ bands is a rather a delicate
quantity\cite{convergeeg}.  In contrast, the LDA underestimates
the spacings between peaks 1 and 2 by $\sim$1.5~eV and between
peaks 2 and 3, by $\sim$0.7.  Moreover, it overestimates the
spectral weight of the first peak by a factor of $\sim$2.  This
result is also significant, because the EELS spectra largely
reflect the O 2$p$ partial DOS.  Without coupling between the Ni
$e_{g}$ level, the itinerant band would adopt a simple
parabolic form; thus the amplitude of first peak is a reflection
of the hybridization between the Ni $e_{g}$ and the itinerant
band.  The fact that \scgw\ gets the correct weight for this peak
establishes that it accurately estimates this coupling, while the
LDA overestimates it by a factor of 2.

Many of the results found here confirm many conclusions drawn in
a model {\em GW} calculation \cite{massidda97}, as well as 
various LDA+$U$
calculations\cite{anisimov93,Dudarev98,bengone00}, both of which
may be viewed as model approaches to the present theory.  Some
significant differences do arise.  The relative positions of
different bands and the energy gaps depend rather sensitively on the
choice of parameters in the model approaches.  For example, in
Ref.~\cite{bengone00}, the O-derived $sp$ conduction-band appears
to fall at 2.8~eV, $\sim$2~eV below the middle of the $e_{g}$
level when $U$ is assumed to be 5 (somewhat lower than the
constrained LDA estimate, $U${}$\sim$8~eV).
Massidda's model $GW$ calculation\cite{massidda97} shows the $sp$
band $\sim$1~eV above the $e_{g}$.

To what extent does the Landau QP picture based on the preceding
\scgw\ results fail to describe the true electronic structure of
MnO and NiO?  We have shown that a great deal is correctly
described, including many details of the valence and conduction
bands.  The main discrepancy is with XPS measurements.  For
optics, the peak in Im($\epsilon$) corresponding to the gap in
NiO is about VBM+4~eV whereas this peak is at VBM+5~eV in the BIS
data.  But Im($\epsilon$) is directly related to the excitonic
process or correlational motion of electron-hole pair. Such a
correlation can shift Im($\epsilon$) downward.  The difference
between the two experiments can be due to this correlation.  So
the poles of the true Green's function (which are reflected in
the DOS) should correspond to the unoccupied $d$ position at
VBM+5~eV as is shown in BIS.  Peak `1' in the \scgw\ DOS falls
slightly higher than experiment, at $\sim$5.8~eV.  If we include
correlation beyond RPA, e.g. inclusion of ladder diagrams,
the band gap in the Green's function may be reduced about 1~eV,
as estimated by Takahashi and Igarashi\cite{takahashib}.  Thus,
it would seem that the RPA explains quite well the important
experimental data, apart from a slight tendency to underestimate
screening of $W$\cite{estw}.  We have not yet attempted to
include excitonic effects, 
so we cannot say to what extent photoemission data can
be explained within the RPA, though estimates in a model context
were reasonably successful\cite{anisimov93}.  Thus, we believe
that the band picture\cite{terakura84} for NiO is a reasonable starting point for
the description of the electronic structure of NiO, much better
than previously thought, and in many respects more appropriate
than the ligand-field picture.

\begin{table}
\caption{Magnetic moments and minimum gaps in MnO and NiO.\label{tab:mm}}
\begin{tabular}{l|rrrr|rcl}
              & \multicolumn{3}{c}{moment} && \multicolumn{3}{c}{bandgap}\\
Compound      & LDA   &\scgw &  Expt       &\ &\ LDA  &\scgw &  Expt   \\
\colrule
MnO           & 4.48  & 4.76 &  4.6        &&     0.78  & 3.5 &  3.9$\pm$0.4 \\
NiO           & 1.28  & 1.72 &  1.9        &&     0.45  & 4.8  &  4.3 \\
 \end{tabular}
 \end{table}

This work was supported by the Office of Naval Research.

\bibliography{pap}

\begin{thebibliography}{22}
\bibitem[{con()}]{convergeeg}
\bibinfo{note}{To achieve this result, it was essential to use a well-converged
  basis set, in particular including local orbitals on the Ni $d$ sites.
  Without the local orbital the $e_{g}$ falls at $\sim$1-2~eV higher energy.}
\bibitem[{est()}]{estw}
\bibinfo{note}{Interestingly, the semiconductors also demonstrate a slight but
  universal tendency to overestimate the bandgaps; see, e.g. Ku's results for
  Ge{\cite{weiku02}}. Not surprisingly, the pair-correlation effect between
  electron-hole is rather larger in a correlated material such as NiO.}
\end{thebibliography}

\end{document}